\documentclass{ws-ijmpb}

\begin{document}

\markboth{Authors' Names}
{Instructions for Typing Manuscripts (Paper's Title)}

\catchline{}{}{}{}{}
%
%%%%%%%%%%%%%%%%%%%%%%%%%%%%%%%%%%%%%%%%%%%%%%%%%%%%%%%%%%%%%%%%%%%%
\title{Topological effects of charge transfer in telomere G-quadruplex:
Mechanism on telomerase activation and inhibition}

%\LaTeX\footnote{For the
%title, try not to use more than 3 lines.
%Typeset the title in 10~pt Times Roman, uppercase and boldface.}

\author{Xin Wang\footnote{Present address: Center for Quantitative Biology, Academy for Advanced Interdisciplinary Studies, Peking University, Beijing, 100871, China}
\ \ and  Shi-Dong Liang
\footnote{Corresponding author Email : stslsd@mail.sysu.edu.cn }}

\address{State Key Laboratory of Optoelectronic Material and Technology, and\\
Guangdong Province Key Laboratory of Display Material and Technology,\\
School of Physics and Engineering, \\
Sun Yat-Sen University, Guangzhou, 510275, People's Republic of China\\
stslsd@mail.sysu.edu.cn}

%\footnote{State completely without abbreviations, the
%affiliation and mailing address, including country. Typeset in
%8~pt Times Italic.}\\%firstauthor
%\footnote{Typeset author e-mail address in single line.} }

%\author{SECOND AUTHOR}
%\address{Group, Laboratory, Address\\
%City, State ZIP/Zone, Country\\
%secondauthor\_id@domain\_name}

\maketitle

\begin{history}
\received{Day Month Year}
\revised{Day Month Year}
%\accepted{(Day Month Year)}
%\comby{(xxxxxxxxxx)}
\end{history}

\begin{abstract}
We explore charge transfer in the telomere G-Quadruplex (TG4) DNA theoretically by
the nonequilibrium Green's function method, and reveal the topological effect of charge transport in TG4 DNA. The consecutive TG4(CTG4) is semiconducting with $0.2\sim0.3eV$ energy gap. Charges transfers favorably in the consecutive TG4, but are trapped in the non-consecutive TG4 (NCTG4). The global conductance is inversely proportional to the local conductance for NCTG4. The topological structure transition from NCTG4 to CTG4 induces abruptly $\sim 3nA$ charge current, which provide a microscopic clue to understand the telomerase activated or inhibited by TG4. Our findings reveal the fundamental property of charge transfer in TG4 and its relationship with the topological structure of TG4.
\end{abstract}


\section{ Introduction}
Charge transfer along double-stranded DNA has attracted much attention among
biomedicine, chemistry and physics communities in the past decade.\cite{Achim,Tapash} As a specialized DNA sequence, telomeres involve many essential physiological processes, DNA damage, cell replication, aging, genetic stability and cancer.\cite{Mergny,Hackett} Human telomeres consist of tandem repeats of the hexanucleotide $(TTAGGG)n$ 5-10kb in length 5'-3' toward the chromosome end, terminating in a single-stranded 3'-overhang of 100-200 bases.\cite{Blackburn} In normal cells, the telomere shortens
with each cell replication. When a critical length is reached, cells undergo apoptosis.\cite{Hurley2} Nevertheless, 80-90\% of cancer cells preserve their telomere length by the activation of telomerase and
thus become abnormal.\cite{Piatyszek} Interestingly, the G-rich sequence
can fold into G-quadruplex (G4) (see in Fig.1), which is a secondary structure
consisting of stacked G-tetrad planes connected by a network of Hoogsteen
hydrogen bonds and stabilized by centre monovalent cations, such as $Na^+$ and
$K^+$. For human telomeric DNA, the repeats of the telomere $(TTAGGG)n$ form the telomere G4 (TG4),
where intramolecular G4 forms three-layer packet (Fig.1),\cite{Dai} which can form several topological structures. It has been found that the formation of TG4 inhibits telomerase activity, which
obstructs tumor immortal mechanism.\cite{Zahler} Hence, TG4 has attracted
extensive studies as an attractive target for cancer therapeutic
intervention.\cite{Mergny}

Moreover, because of DNA self-assembly properties, its paring specificity
and conformational flexibility offer great potential for the rational design
of DNA-based nanostructure and nano electronics.\cite{Cuniberti} Different topological structures of G4 DNA
provide more possibilities for nano devices.

However, there are divergent opinons on charge transfer in the G-rich DNA sequences.
Sugiyama et. al. believe that the G base is favor for charge transfer
because the ionization potential of G base is lowest among the nucleobases.\cite{Sugiyama}
The numerical study also shows G4 DNA favoring charge transport based on a
simplified mono-G tight-binding G4 model.\cite{Guo} In fact, charge transfer in G4
could be quite different from normal DNA due to the cation effect as a new
hydrogen bond in G4 DNA. Barton's group found experimentally that G4 has
great trapping potency on charges by chemical florescence method.\cite{Delaney}
Actually, the problem of charge transfer in G4 is still open because of the complicated topological
structures of G4. The full understanding of charge transfer in G4 requires comparing different
topological structures of G4 DNA. However, it is not easy to compare
functions of different topological structures of G4 by biomedical methods.
Physical method can provide a microscopic insight to understand the
mechanism of the G4 functions in physiological process and charge transfer
for nano devices.

In this paper we will study charge transfer in intramolecular TG4
by the effective tight-binding model with nonequilibrium Green's
function (NEGF) method. Investigating the current-voltage (I-V) characteristics of two predominant
topological structures of TG4 in vivo, we find novel charge transfer
properties of TG4 beyond the classical charge transfer law, which provide basic properties for
designing G4-based nano devices and inspire a physical mechanism on the telomerase activation
inhibited by TG4.\cite{Zahler}

\section{General G4 model}
The TG4 DNA can be viewed as comprising of four parallel $\pi$ stacks for
conducting charge channels through the superposition of $\pi$ orbitals along DNA molecules, which is
observed by the NMR spectroscopy and x-ray crystallography.\cite{Feigon} We consider the two ends of
G4 DNA connected with two semi-infinite one-dimensional electrodes. The Hamiltonian of G4-DNA model can be written as

\begin{equation}
H=H_{G_4}+H_{L,R}+H_{G_{4}-LR}+H_{env}
\label{H}
\end{equation}%
The charge channels in the backbone of G4 DNA is very small and the $\pi$ orbit channel dominates charge transfer in G4 DNA.\cite{Cuniberti} Thus, the effective tight-binding Hamiltonian of G4 DNA $H_{G_{4}}$ can be expressed as
\begin{eqnarray}
H_{G_{4}}&=& \sum\limits_{n=1,k\in G_{4}}^{N}\left[ \varepsilon
_{nk}c_{n}^{\dagger}c_{n}-t_{n,n+1}\left( c_{n}^{\dagger}c_{n+1}+c_{n+1}^{\dagger}c_{n}\right)\right]\\\nonumber
&-& \sum\limits_{\langle k,\ell \rangle \in G_{4}}t_{m}\left(
c_{k}^{\dagger}c_{l}+c_{l}^{\dagger}c_{k}\right)
\label{HG4}
\end{eqnarray}
where $\varepsilon _{nk}=\varepsilon _{n}+\frac{e^{2}}{4\pi \varepsilon
_{r}R_{0}}\delta _{nk}$ is the on-site energy of each bases,\cite{Sugiyama} where
the second term is  the ion effect in G4 packet, where $R_{0}=1.99nm$ is the distance between the ion
and $\pi$ electrons, and $\varepsilon _{r}\approx 2\varepsilon _{0}$ is the effective dielectric constant.\cite{Neidle2} The second term in Eq.(\ref{HG4}) describes the $\pi$ electron hoppings between Gs in G4 packet.
The $t_{n,n+1}$ is the nearest $\pi $ electron hopping parameters listed in table I.\cite{Sugiyama,Senthilkumar}
The notation 5'-XY-3' indicates the direction along the DNA strand (see,Fig.1(b) \cite{Endres}.
The $t_{m}$ describes the hopping amplitude between Gs in G4
packet. The $H_{L,R}$ in Eq.(\ref{H}) is the electrodes at the left and right ends of G4 DNA,
\begin{equation}
H_{L,R}=\sum\limits_{n\in L,R}\left[ \varepsilon
_{c}c_{n}^{\dagger}c_{n}-t_{c}\left( c_{n}^{\dagger}c_{n+1}+c_{n+1}^{\dagger}c_{n}\right) %
\right].
\label{HLR}
\end{equation}
The $H_{G_{4}-LR}$ in Eq.(\ref{H}) is the couplings between the G4 DNA and electrodes,
\begin{equation}
H_{G_{4}-LR}=-t_{L}\left( c_{0}^{\dagger}c_{1}+c_{1}^{\dagger}c_{0}\right) -t_{R}\left(c_{N}^{\dagger}c_{N+1}+c_{N+1}^{\dagger}c_{N}\right).
\label{HG4LR}
\end{equation}
The $c_{k}^{\dagger}(c_{k})$ is the creation (annihilation) operator of
electron or hole at k sites. The $\varepsilon _{L(R)}$ ( $t_{L(R)}$ ) is the
on-site energy (hopping parameter) in the electrode,
respectively. To minimize the contact effect, we use the optimal injection
condition $\sqrt{t_{\alpha }t_{L(R)}}=t_{L(R)},$and the strong coupling in the
electrode $t_{L(R)}=1eV$,\cite{Macia} where $\alpha $ denotes A, C, G, and T.

\begin{table}
\tbl{The on-site energy and the $\pi$ electron hopping parameter between nearest bases.}
%\begin{indented}\item[]
{\begin{tabular}{ccccc}
\hline\hline
&  &   5'-XY-3'(eV) &   &   \\\hline
X/Y & G & C & A & T  \\\hline
G & 0.119 & -0.295 & -0.186 &  0.334  \\
C & 0.026 & 0.042 & -0.008 &  -0.161  \\
A & -0.013& 0.091 & -0.038 &  -0.157  \\
T & 0.044 & -0.066& -0.068 &  0.180  \\\hline
$\varepsilon_{n}$ & 7.75 & 8.87 & 8.24 & 9.14\\
\hline\hline
\end{tabular}}
%\end{indented}
\label{Table 1}
\end{table}

Since realistic DNA molecules are under physiological conditions, we consider $TG_{4}$ DNA having water and ion environment. These water and ion environments of $TG_{4}$ provides a stochastic electric field background to influence the $TG_{4}$ structure as a stochastic perturbation.
We introduce the $H_{env}$ describing these environment effects by the stochastic
fluctuation of the $G_4$ structure. Thus, the $H_{env}$ has the same form to $H_{G_4}$, but
the parameters are $\varepsilon^{env}_{n}=0.5\lambda r$ and $t^{env}_{n,n+1}=0.5t_{n,n+1}\lambda r$,
where $\lambda$ describes the stochastic strength and $r\in (-1,1)$ randomly.\cite{Giese,Elstner}

It should be remarked that we introduce the electrode such that we can
calculate the I-V characteristic, which can be compared directly
with experiments. For the case of G4 DNA in vivo,
suppose that the physiological environment provides an electric potential bias to G4 DNA. Thus,
the situation of G4 DNA in vivo maps similarly to the case we consider.

Using NEGF Method, the current can be expressed in terms of \cite{Viljas}
\begin{equation}
I=\frac{e}{2\hbar }\int \frac{dE}{2\pi }Tr[G^{r}\Gamma _{R}G^{a}\Gamma_{L}](f_{L}-f_{R})
\label{I}
\end{equation}
where $G^{r(a)}=[E-H_{G_{4}}-\sum_{L}^{r(a)}-\sum_{R}^{r(a)}]^{-1}$ is the
retard (advanced) Green's Function; the $\Gamma_{L(R)}=i(\sum_{L(R)}^{R}-\sum_{L(R)}^{A})$ describes
the level-width function, where $\sum_{L(R)}^{r(a)}$ is the retard (advanced) self-energy of electrodes.
The $f_{L(R)}$ is the Fermi function of the left (right) electrodes.

\section{TG4s and their classifications}

In general, the telomere sequence can be folded to many types of TG4
DNA. The human telomere $(TTAGGG)_n$ has been found to be folded to two topological
structures (TP1 and TP2) in vivo shown in Fig.1.\cite{Neidle2}
The TG4 packet as an basic element can be constructed to different configurations of TG4 DNA,
which can be classified into three classes, the consecutive mono(hybrid)-TG4(CM(H)TG4) DNA,
and non-consecutive mono(hybrid) TG4 (NCM(H)TG4) DNA. Moreover, we may compare the TG4 packet with and without ion.
For convenience, we label the TG4 DNA by $sn(ijk)i$ or $sn(ijk)a$, where sn
labels the telomere sequence $(TTAGGG)_n$; the $(ijk)$ labels the TG4 topological structure and the last $i$ and $a$ labels ion and absence of ion. For example we define $s16\equiv TTA(GGGTTA)_{16}$ and when $s16$ is folded to two TP1s and TP2s with ion. We label it by $s16(1122)i$ that is CHTG4. When $s16$ is folded to one TP1 and two TP2 without ion. We label it as $s16(1022)a$ that is NCHTG4, where 0 labels the sequence of non-TG4 packet.

\begin{figure}
\begin{center}
\includegraphics[scale=1.3]{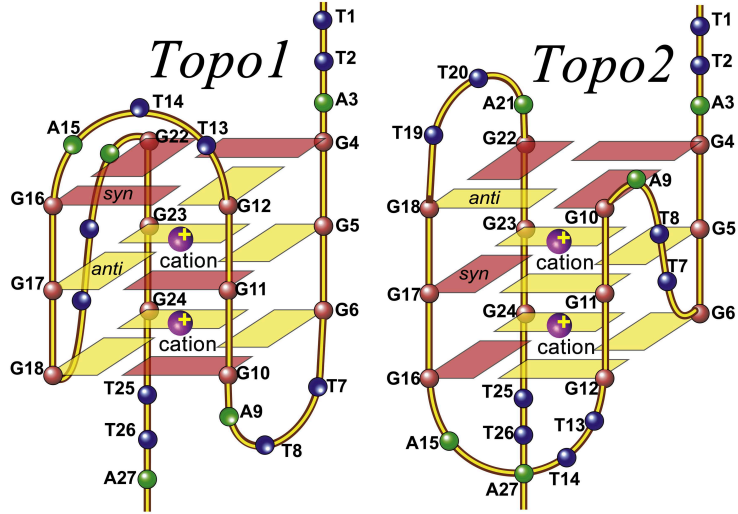}% Here is how to import EPS art
%\special{epsfile=Fig1.eps voffset=0 hoffset=0
%vscale=100 hscale=100 angle=0}
\end{center}
\caption{\label{label}(Color online) The sketch of two topological structures of TG4 DNA in human vivo.}
\end{figure}

\section{Topological effect of charge transfer}
In order to capture the basic physics of the charge transfer in $TG4$, we first turn off the environment effects $H_{env}=0$. For CMTG4, $s4\equiv TTA(GGGTTA)_4$, it can be folded to
only one TG4 packet. We compare the I-V characteristic of the s4 telomere
chain and two topological structures (TP1 and TP2) of TG4 with and without
ion, namely $s4(1)i$, $s4(2)i$, $s4(1)a$, and $s4(2)a$, shown in Fig.2 (a).
The s4 chain is of little conductance compared with CMTG4. The $s4(1)i(a)$ and $s4(2)i(a)$ show semiconducting with the energy gap 0.3eV. The currents saturate at around 1V due to only finite states for molecule
systems. The saturated currents of $s4(1)i$ and $s4(2)i$ reach
6nA and 1nA respectively, which is much higher than that of $s4(1)a$, and $s4(2)a$
(0.06nA). That means that the ion in TG4 enhances charge transfer. For
longer telomere sequences, $s12\equiv TTA(GGGTTA)_{12}$, and its CMTG4,
the basic behaviors of the I-V characteristics are similar to that of
the s4 case, but the energy gaps for the non-ion cases
are very small. The saturated currents of $s4(2)a$ is
larger than that of $s4(1)a$, but reverse for the $s12$ case, which can been seen in Fig.2 (b).

\begin{figure}
\begin{center}
\includegraphics[scale=1.7]{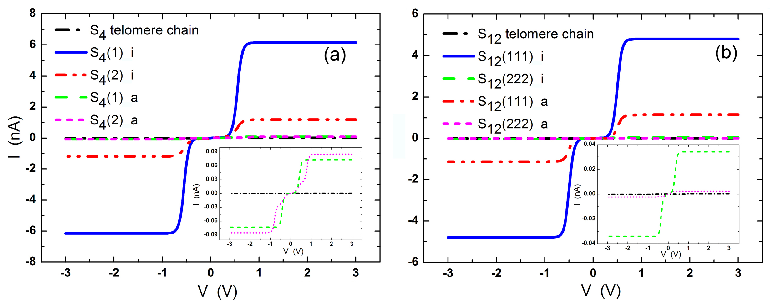}% Here is how to import EPS art
%\special{epsfile=Fig2.eps voffset=0 hoffset=0
%vscale=130 hscale=130 angle=0}
\end{center}
\caption{\label{label}(Color online) The I-V characteristic of CMTG4 DNA in (a) for the $s4(j)i(a)$,
in (b) for the $s12(jjj)i(a)$, where $j=1,2$ means TP1 or TP2.}
\end{figure}

For CHTG4 the basic physical behavior of I-V
characteristics is same to that of CMTG4. They are also
semiconductor with $\sim0.3eV$ energy gap and the saturated
currents of the case with ion are much larger than that of the cases without
ion. The saturated currents depend on the configurations of the topological structures of TG4 shown in Fig.3.

\begin{figure}
\begin{center}
\includegraphics[scale=1.7]{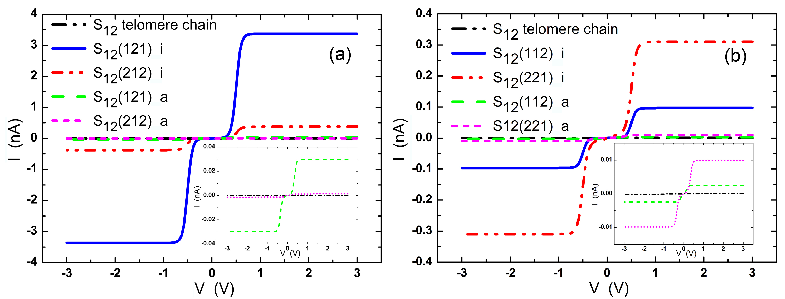}% Here is how to import EPS art
\end{center}
\caption{\label{label}(Color online) The I-V characteristic of CHTG4 DNA in (a) for the $s12(jkj)i(a)$, and
in (b) for the $s12(jjk)i(a)$,where $j=1,2$ means TP1 or TP2.}
\end{figure}

For NCHTG4 in Fig.4, the saturated currents of NCHTG4 with and without ion become very small $0.002\sim 0.2pA$ and actually vanish for some cases, which is quite different from CMTG4 and CHTG4. The saturated current of the telomere chain is larger than that of NCM(H)TG4. It implies that TG4 in NCM(H)TG4 can trap charges, which is consistent with the experimental results.\cite{Delaney}

\begin{figure}
\begin{center}
\includegraphics[scale=1.7]{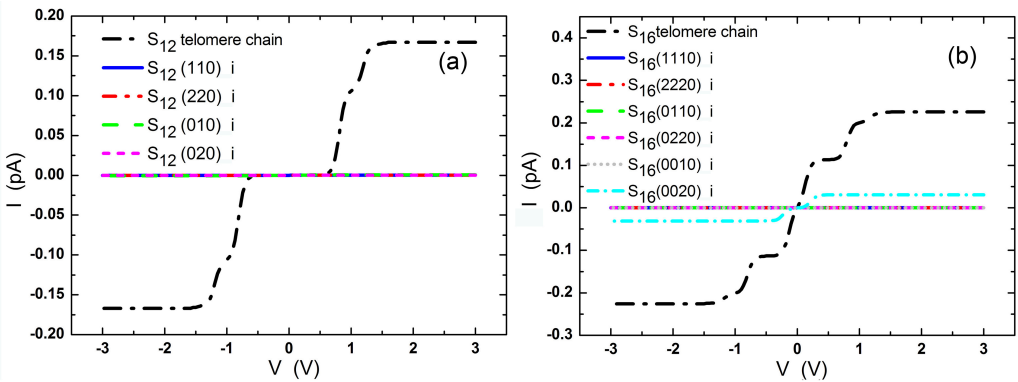}% Here is how to import EPS art
\end{center}
\caption{\label{label}(Color online) The I-V characteristic of NCMTG4 DNA in (a) for the $s12(jj0)i(a)$
and $s12(0j0)i(a)$, and in (b) for the $s16(jjk0)i(a)$, $s16(0jj0)i(a)$ and  $s16(00j0)i(a)$.}
\end{figure}

Charge transport in NCM(H)TG4 exhibits novel topological properties.
In common sense of charge transport in classical and quantum systems,
the global conductance of the system is proportional to the local conductance of the system. However, for NCH(H)TG4, the global conductance is inversely proportional to the local conductance. The conductance of TG4-DNA system may be regarded as to be proportional to the saturated current. We find the conductance $G_{TG4}>G_{s4}$ from results in Figs.2 and 3. When the telomere chain $s16$ forms NCMTG4, $s16(1000)i$, the conductance becomes $G_{s16(1000)i}<G_{s16}$ shown in Fig.4. It implies that the global conductance of NCM(H)TG4 is inversely proportional to the local conductance even though the local conductance $G_{TG4}>G_{s4}$. In other words, charge transfer in NCM(H)TG4 exhibits nonlinearity.
The topology of TG4 induces the anomalous charge transport.
We can understand these novel charge transfer properties in NCM(H)TG4 from quantum mechanics. Physically,
mobile charges form Bloch wave in a periodic or quasi-periodic structures of CM(H)TG4, while mobile charges in NCM(H)TG4 are trapped due to Anderson's localization effect in disorder chain. That is why charges transfer favorably in the CM(H)TG4. Actually, DNA sequence may be regarded as a quantum system. Charge transfer in DNA
sequence has been found exhibiting some quantum nature.\cite{Tapash}
The anomalous conductance in NCM(H)TG4 we find is a novel quantum property of charge transfer induced by topology of TG4 DNA.

\section{Topological structure transition and mechanism on telomerase activation and inhibition}
The charge transfer properties of TG4 DNA can give a physical clue to understand TG4 how to activate or inhibit telomerase. It has been found that $80\%\sim90\%$ cancer cells obtain their immortality by activating telomerase to preserve the telomere length.\cite{Piatyszek} The telomerase activation or inhibition is related to TG4 DNA.\cite{Piatyszek}  The challenging problem is how to activate or inhibit the telomerase.\cite{Kim,Hanahan}
Suppose that the ligands of the telomerase offer an electric bias on the telomere DNA to form a circuit inducing charge current. The topological structure transition from NCM(H)TG4 to CM(H)TG4 generates abruptly the charge current $3nA$ from $0.002pA$ (see Figs. 2 and 4). This current as a threshold current can activate or inhibit telomease. This scenario can provide a clue to understand TG4 blockading the cancer cell immortal pathways and eventually perishing tumors in the biomedical experimental observation.\cite{Zahler} This finding inspires a further challenging problem for biomedicine how to realize or control the topological structure transition between NCM(H)TG4 and CM(H)TG4.

\begin{figure}
\begin{center}
\includegraphics[scale=1.3]{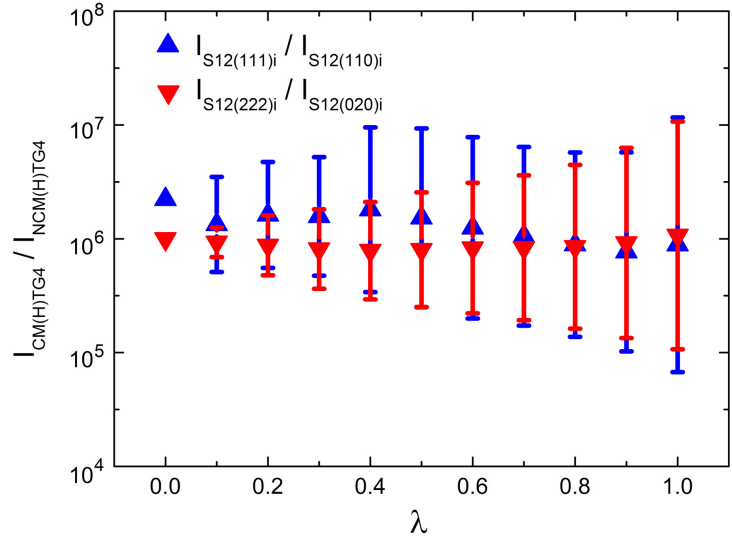}% Here is how to import EPS art
\end{center}
\caption{\label{label}(Color online) The ratio of the currents of CM and NCM TG4s at $V=1V$. We take 20 samples to give the average and the standard errors. The values of the ratios still remain $6$-order differences against the environment effects.}
\end{figure}

\section{Discussion}
It should be remarked that the ratio of the saturated currents between CM(H)TG4 and NCM(H)TG4 has $6$ orders.
The solvent leads only $0.1\sim 0.5$eV on-site energy fluctuation to DNA bases, which cannot lead to higher G4 structures.\cite{Elstner} We investigate the environment effects in Fig. 5 for two cases. When the environment fluctuation is small $\lambda<0.5$, the error bar of the ratio of the currents for the $S12(111)i/S12(110)i$ is larger than that of the $S12(222)i/S12(202)i$, which implies that the current fluctuation depends on the detail configurations of the TG4 in the small environment fluctuation. When the environment fluctuation becomes larger, $\lambda>0.6$, the current fluctuation trends to be insensitive to the detail configurations of TG4. More importantly, we do see that the $6$-order current difference induced by the topological structure transition from NCM to CM TG4s is robust against the environment fluctuation. Moreover, it has been found that two basic mechanisms of charge transfer in DNA chain, coherent tunneling (hopping), and thermal diffusive hopping.\cite{Giese2} The coherent tunneling or hopping occurs betwen $G/C$ pairs and the thermal hopping occurs in more than $3$ consecutive $A/T$ pairs.\cite{Giese2} For TG4, $(TTAGGG)n$, the coherent hopping dominates the charge transfer due to only $3$ consecutive $A/T$ pairs in TG4. Thus, the coherent transport method is valid for TG4. In addition, the energy gap at Fermi energy is about $0.2\sim0.3eV$ for TG4, which is $1$ order higher than the influence of temperature fluctuation in room temperature, and the charge transfer in TG4 is not sensitive to temperature fluctuation because the coherent hopping dominates the charge transfer in TG4 instead of the thermal diffusive hopping. Namely, temperature fluctuation in room temperature is not sensitive for our results. Therefore, in room temperature all our conclusions are invariant.

\section{Conclusion}
In summary, we reveal novel charge transport properties in TG4 DNA. The CM(H)TG4 is semiconducting
with $0.2\sim0.3eV$ energy gap. The conductances of different configurations of TG4 DNA follow the inequalites, $G_{CM(H)TG4}>G_{TC}>G_{NCM(H)TG4}$, where $TC$ means telomere chain. The conductances depend on the topological structures and configurations of TG4. The cation in TG4 enhances charge transfer. The TG4 packet in NCM(H)TG4 suppresses charge transfer, which agrees with the experimental results.\cite{Delaney} The topological structure transition between NCM(H)TG4 and CM(H)TG4 induces a threshold charge current to activate or inhibit telomease. Our findings not only reveal novel charge transfer properties in TG4 DNA that can offer many opportunities for DNA-based electronics, but also provide physical insights and hints to understand the telomerase activation inhibited by TG4.

The biomedical investigation found that tumor growth is related to the telomerase activation and inhibition. The telomerase activation and inhibition is related to TG4. \cite{Piatyszek} It implies strongly that states of TG4 could be related tumor growth even cancer. However, the biomedical investigation cannot find the working mechanism of these three factors, TG4, tumor growth and telomerase activation and inhibition. Our results provide an understanding of the working mechanism of the telomerase activation and inhibition.
We found that the topological structure transition from NCM(H)TG4 to CM(H)TG4 generates abruptly the charge current $3nA$ from $0.002pA$ (6 order difference) when the ligands of the telomerase offer an electric bias on the telomere DNA. It strongly implies that this abrupt current as a threshold current can activate or inhibit telomease. Namely, if we can control the topological structure transition from NCM(H)TG4 to CM(H)TG4 we can control the tumor growth. This scenario can provide a hint to understand TG4 blockading the cancer cell immortal pathways and eventually perishing tumors in the biomedical experimental observation.\cite{Zahler}
Consequently, the further challenging issue is how to tune the topological structure transition of TG4 by biomedical method, which is expected to further study.

These novel charge transfer properties in TG4 DNA give some fundamental relationships between topology, anomalous charge transport and biomedical function.

\section*{Acknowledgements}
The authors acknowledge the financial supports of the projects
from the Elite Student Program from National Education Department, and the Fundamental Research Funds for the Central Universities.

%\appendix{Heading of Appendix}
%\subsubappendix{Sub-subappendix}
%\appendix{Another Appendix}

%\begin{figure}[bt]
%\centerline{\psfig{file=ijmpbf1.eps,width=3.65in}}
%\vspace*{8pt}
%\caption{This is the caption for the figure. If the caption is
%less than one line then it is centered. Long captions are
%justified to the full text width.}
%\end{figure}

\end{document}